\documentclass[a4paper,11pt]{article}

\usepackage{pos}
\usepackage{lipsum} %package to generate placeholder text in the following

\title{Sentinel of the extraordinary: the IceCube alert system for neutrino flares}

\ShortTitle{IceCube alert system for neutrino flares}

\author{The IceCube Collaboration \\{\normalsize \normalfont(a complete list of authors can be found at the end of the proceedings)}\\}

\emailAdd{caterina.boscolomeneguolo@studenti.unipd.it}
\emailAdd{elisa.bernardini@unipd.it}
\emailAdd{sarahlouise.mancina@unipd.it}

\abstract{

The IceCube Neutrino Observatory has the invaluable capability of continuously monitoring the whole sky. 
This has affirmed the role of IceCube as a sentinel, providing real-time alerts to the astrophysical community on the detection of high-energy neutrinos and neutrino flares from a variety of astrophysical sources. As a response to the IceCube alerts, different observatories can join forces in the multi-messenger observation of transient events and the characterisation of their astrophysical sources. 
The 2017 breakthrough identification of blazar TXS 0506+056 as the source of high-energy neutrinos and UHE gamma rays was proof of this strategy.
The Gamma-ray Follow-Up (GFU) is the IceCube program for identifying high-energy muon neutrino single events, as well as outstanding neutrino flares from relevant sources and the whole wide universe. 
While the identification of single high-energy neutrinos is shared on public alert distribution networks, partner Imaging Air Cherenkov Telescopes are sent low-latency alerts following the detection of neutrino flares, for which they have dedicated follow-up programs. 
I will present an overview of the GFU platform together with new results from the analysis of recorded neutrino flares, after a dozen years of GFU operation and hundreds of alerts being sent.
\vspace{4mm}
{\bfseries Corresponding authors:}
Caterina Boscolo Meneguolo$^{1*}$, Elisa Bernardini$^{1}$, Sarah Mancina$^{1}$\\
{$^{1}$ \itshape Università degli Studi di Padova}\\[4mm]
$^*$ Presenter

\ConferenceLogo{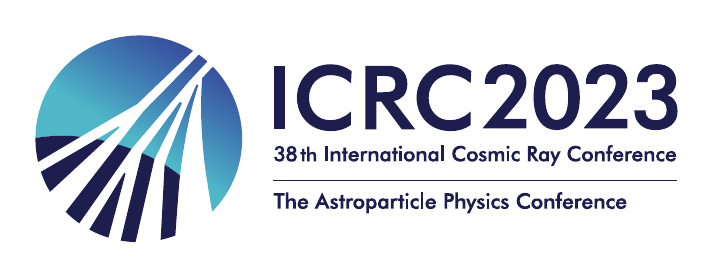}

\FullConference{The 38th International Cosmic Ray Conference (ICRC2023)\\ 26 July -- 3 August, 2023\\ Nagoya, Japan}
}

\begin{document}

\maketitle

\section{Introduction}\label{intro}
It has long been hypothesized that astrophysical sources of neutrinos should also produce gamma-ray emission via pion decay, which can be observed if the gamma-rays are not obscured by matter or attenuated by background light on their journey to Earth.
The signal of astrophysical neutrinos seen by IceCube is obscured by a background of muons and neutrinos produced by cosmic ray air showers in the atmosphere.
These backgrounds are a nuisance when looking for statistically significant clustering of neutrino events in the sky to identify the astrophysical sources of neutrinos.
Many candidate astrophysical neutrino sources have been known to have variability in various wavelengths of light.
Therefore, the relatively time-uniform atmospheric backgrounds can be reduced by using time-dependent analyses which search for flares in neutrinos.

IceCube has an uptime of around 99\% and the ability to observe the full sky at all times during its operation.
On the contrary, Imaging Air Cherenkov Telescopes (IACTs), the most sensitive instruments to very-high-energy gamma-rays ($>$ 100 GeV), can only operate at night and have a limited field of view.
A system that triggers IACT follow-up of neutrino flares detected in IceCube data is therefore the ideal setup for collaboration between the two types of experiments.
Telescopes that could potentially follow-up a neutrino flare in other wavelengths of light are also limited by these same constraints.
The advantages of this multi-messenger approach was exemplified in the analysis of neutrino emission from TXS 0506+056~\cite{IceCube:TXS0506MMA}.
In this instance, IACTs and other telescopes were alerted by a single high energy neutrino (singlet) from IceCube.
An archival analysis of the IceCube data revealed a neutrino flare comprised of several events with energies below the singlet alert threshold had occurred in 2014~\cite{IceCube:TXS0506NuFlare}.
Therefore, looking for time-dependent excesses within the IceCube data can provide an additional avenue for alerting telescopes to locations of potentially interesting sources.
Presented here is the alert stream for IceCube neutrino flares developed for alerting IACT telescopes of potential targets of opportunity; hence, we refer to this analysis as the Gamma-Ray Follow-up alert stream.

The current version of the Gamma-Ray Follow-Up (GFU) alert stream was put into operation in 2019~\cite{Kintscher2020}, but previous versions of the Neutrino Triggered Target of Opportunity, program have been running since 2006 with IceCube's predecessor AMANDA~\cite{IceCube:GFU2016}.
The alert streams produced consist of an analysis which looks for time-evolving neutrino flare candidates from the direction of known nearby blazars and an analysis which looks for significant neutrino flare candidates across the whole sky.
In Section~\ref{gfu_es} we provide a brief description of the event selection used to gather neutrino candidates from the IceCube data.
Section~\ref{gfu_alerts} covers the analysis construction for identifying statistically significant neutrino flares to send as alerts.
We also discuss preliminary results from an offline version of the alert stream run on the archival IceCube data in Section~\ref{results}.
Finally, Section~\ref{con} provides conclusions and a discussion on the future outlook and potential developments for this program.
Currently the alerts are shared with the CTA LST-1, H.E.S.S., MAGIC, and VERITAS collaborations.
A report on the gamma-ray follow-ups to these alerts can be found in References~\cite{IceCube:2023VHEGFU, HESS:GFUFollowUp}.

\section{Event Selection}\label{gfu_es}
\begin{figure}
    \centering
    \includegraphics[width=.85\textwidth]{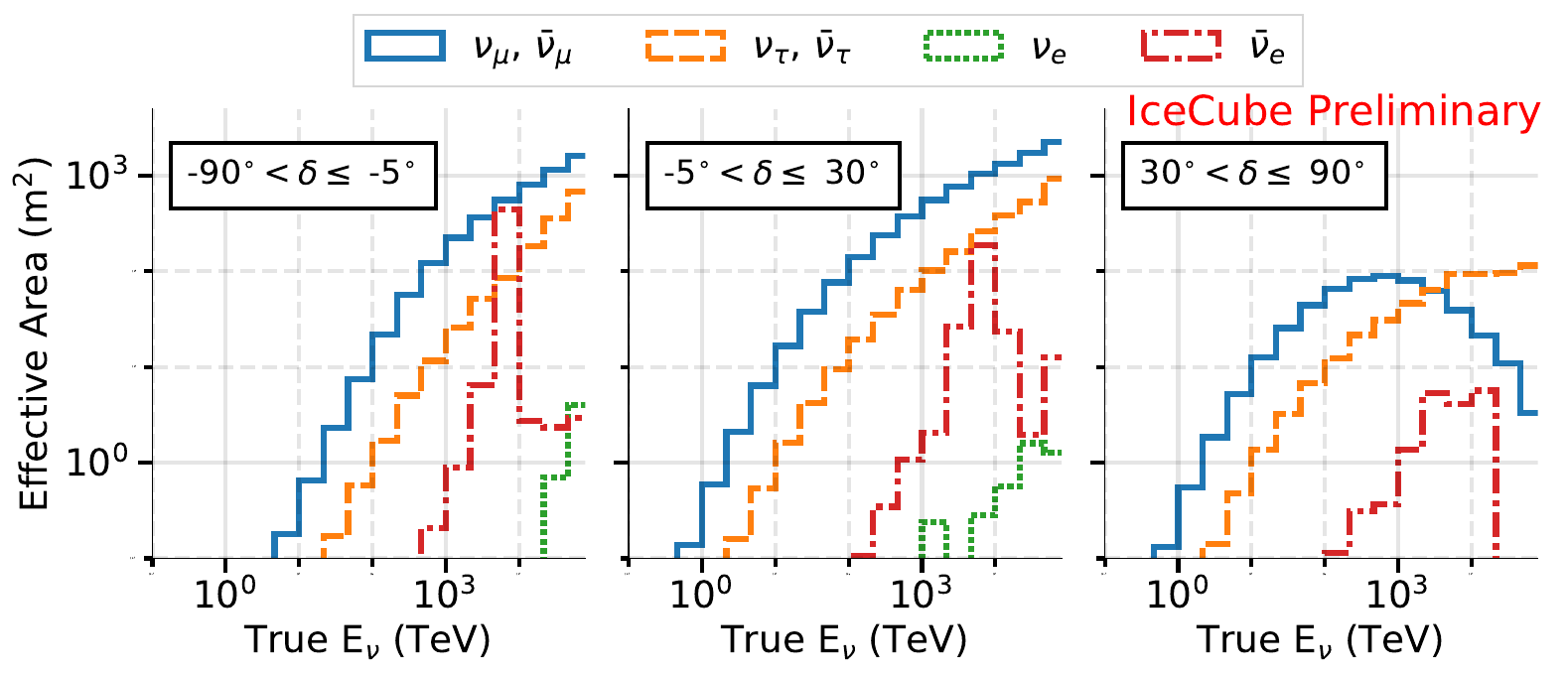}
    \caption{Effective area for three regions of the sky for the IceCube Gamma-Ray Follow-Up (GFU) selection broken into different types of neutrinos. In the southern sky (left), the event selection requires events to have a higher energy in order to reject tracks from the dominating atmospheric muon background.
    The horizon area (middle) shows the best effective area performance.
    In the northern sky (right) a decrease in the effective area is see at the highest energies due to the absorption of neutrino by the Earth since the Earth becomes more opaque to neutrinos at these energies.
    The peak seen in the electron anti-neutrino effective area is due to the Glashow effect which creates a resonance peak around 6.3 PeV in the cross-section.}
    \label{fig:eff_a}
\end{figure}
The Gamma-Ray Follow-Up (GFU) event selection runs in realtime at the South Pole on the IceCube data, and the selected events are sent north for prompt analysis.
The event selection is optimized for time-dependent neutrino source searches by selecting muon track events and prioritizing signal efficiency over background rejection.
Muon tracks, which can come from charged-current muon neutrino interactions, provide the best pointing resolution out of the possible neutrino event morphologies due to the long and straight path the muon produces as it travels through the detector.
For time-dependent analyses, the additional parameter of time naturally reduces the background contamination; hence, more background events from atmospheric neutrinos and muons are allowable in exchange for increased signal capture.
However, in the southern sky, the rate of atmospheric muons greatly overshadows both the astrophysical muon neutrino signal and the atmospheric neutrino background.
Therefore, two independent boosted decision trees (BDTs) are used for up-going events (from the northern hemisphere) and down-going events (from the southern hemisphere), resulting in a higher average event energy in the southern sky sample than that in the north.
The differences in the effective area for the GFU sample in different regions of the sky can be seen in Figure~\ref{fig:eff_a}.
The GFU sample was initially created for sending neutrino flare alerts to IACT telescopes~\cite{IceCube:GFU2016}, but is now used in a variety of time-dependent and realtime analyses~\cite{IceCube:FRA2020, IceCube:2023icecat1, IceCube:GWwithGFU, IceCube:2023OFU}.

\section{Neutrino Flare Alerts}\label{gfu_alerts}
To search for significant clusters of events in time at a given location in the sky, we run the following statistical test~\cite{Kintscher2020}. 
The clustering algorithm is triggered by events which pass a signal-over-background threshold.
The signal-over-background ratio is calculated for each neutrino candidate using probability distribution functions, which account for the event's spatial proximity to the test location and reconstructed energy.
Events are required to have a signal over background ratio greater than 1 to count as a trigger event.
Once triggered, the algorithm looks at previous event triggers for that location that arrived within the previous 180 days to generate time windows with the current event trigger.
For each time window, we calculate a test statistic (TS) using a maximum unbinned likelihood fitting the source location for the number of signal neutrinos, $n_s$, and a single power law spectral index, $\gamma$, with all of the events that fall within the time window~\cite{BRAUN:MLE}.
The time window which results in the maximum TS is chosen as the best-fit value and is compared to a TS distribution derived from background trials to compute a pre-trial p-value.
If the significance of the pre-trial p-value is greater than a given threshold, IceCube sends an alert to the relevant IACTs.

\begin{figure}
    \centering
    \includegraphics[width=0.55\textwidth]{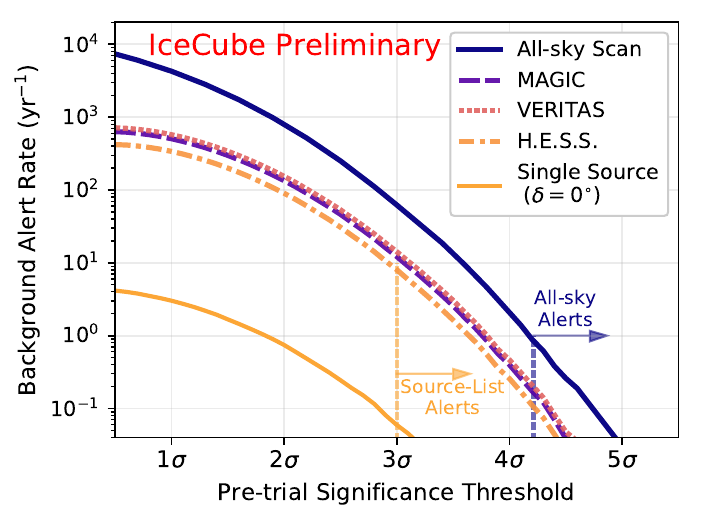}
    \caption{The expected alert rate derived from background trials as a function of the chosen significance threshold. 
    The solid dark blue line shows the rate for the all-sky scan alerts with a dashed line showing the current threshold in operation, which was chosen by selecting where the background rate was equal to one false alarm per year.
    The solid light orange line shows the false alert rate for a signal source located at a declination of 0° with a dashed line representing the current 3$\sigma$ threshold in operation.
    Each IACT has a catalog of sources, here we sum up the background alert rates expected for each source declination to produce a total false alarm rate for each source list.}
    \label{fig:far}
\end{figure}
Using this approach, two types of alerts are provided: source-list alerts and all-sky alerts. 
For the source-list alerts, lists of nearby blazars from the Fermi 3FGL and 3FHL catalogs were produced for each IACT based on the declination range of the experiment~\cite{Kintscher2020}.
Sources were required to have a redshift measurement of z $\leq$ 1, show variability in gamma-rays, and be relatively bright in 100 GeV gamma-rays~\cite{Kintscher2020}.
The selection process resulted in a source list of 179 sources for MAGIC, 190 sources for VERITAS, and 139 sources for H.E.S.S~\cite{IceCube:2023VHEGFU}.
CTA LST-1 has also begun receiving alerts using the same source list as MAGIC.
The GFU alerts search for emissions from the directions of these sources in realtime as GFU events are flagged by the computing system at the South Pole and promptly sent north.
The source-list alert trigger threshold was chosen to be the TS value which gives a pre-trial p-value equivalent to a 3$\sigma$ signficance.
The false alarm rate as a function of the pre-trial p-value threshold is shown in Figure~\ref{fig:far} where the vertical dashed light-orange line represents the current 3$\sigma$ threshold in operation.
Summing the false alarm rates for all sources in each IACT catalog we find that this threshold results in on the order of 10 false alerts from background sent to the IACTs per year.

For the all-sky alerts, the whole sky is monitored for a potential neutrino flare signal.
The advantage of this construction is its model independence; however, it must contend with the statistics due to the increase in trials. 
For each event that passes the online event selection, the algorithm runs on pixels generated by ``healpy"~\cite{Healpy1, Healpy2} within a radius of 2° of the new event's location that have a signal-over-background ratio greater than one~\cite{Kintscher2020}.
The scan begins with a coarse grid search to identify interesting pixels which it will rerun the algorithm on with finer and finer pixels until the target pixel size is reached~\cite{Kintscher2020}.
The pre-trial p-value threshold for sending alerts was chosen to be 4.2$\sigma$ where the false alarm rate reaches about one background alert per year (Figure~\ref{fig:far}).
A similar all-sky alert stream in development that looks for doublets or triplets of GFU events that arrive from similar directions within 30 days of each other is discussed in Reference~\cite{IceCube:2023OFU}.

In order to limit the number of alerts sent for each source, a muting system was implemented.
If a trigger crosses the TS threshold, an alert is sent and the source (or sky location) is then muted, meaning all triggers that occur after the initial alert are ignored until the significance drops below the TS threshold.
This muting procedure avoids overloading the communications with our telescope partners; however, it obscures information about potential increases of the flare significance in time.
Therefore, an offline analysis was performed to look for flares which increased beyond the initial alert report, as well as to look for flares which may have occurred in the archival data before the current stream was operating.
The results for the offline search are given in the following section, Section~\ref{results}.

\section{Offline Analysis Results}\label{results}
The offline flare analyses unveiled the most significant flares in 11.5 years of IceCube data for the source-list analysis and the all-sky analysis.
A post-trials correction for these flares could then be calculated for each analysis.
The post-trials correction must account for each source in the source lists, or pixel for the all-sky alerts, tested as well as exposure time since the number of trials triggered increases as more events are collected from the IceCube data.

\begin{figure}
    \centering
    \includegraphics[width=0.85\textwidth]{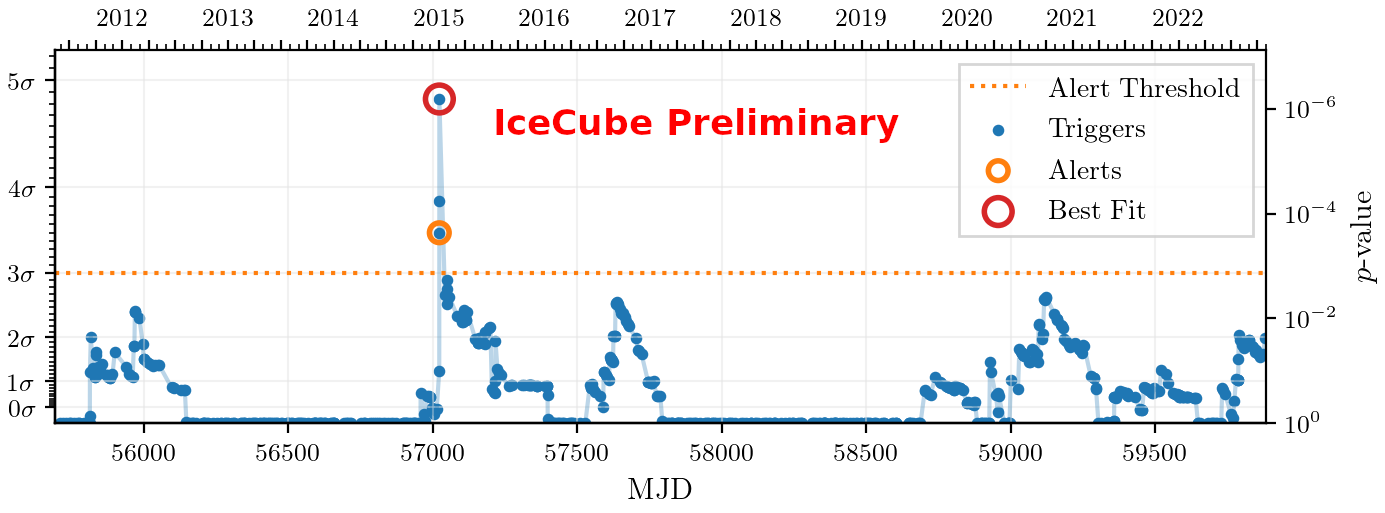}
    \caption{The pre-trial significances of the best-fit flare time-window as a function of the event trigger time for the source 1ES 0347-121 over 11.5 years of data. 
    The orange circled point represents where an alert would have been sent if the alerts had been active in 2014/15.
    The red circle indicates the peak in significance found for the source which after trial corrections drops to 1.81$\sigma$.}
    \label{fig:source_hotspot}
\end{figure}
For the source-list analysis, the best fit flare was found from the direction of 1ES 0347-121.
This source lies in the southern sky at a declination of -11.98° where events tend to be higher in energy.
The flare had a pre-trial significance of 4.84$\sigma$ and, after accounting for the trials factor, the post-trials significance was 1.81$\sigma$.
The time window between trigger events that resulted in the best-fit flare was 6.9 hours with the algorithm returning a best fit of 3.93 signal events.
The significance as a function of trigger times over the 11.5 years of IceCube data for 1ES 347-121 is shown in Figure~\ref{fig:source_hotspot}.

\begin{figure}
    \centering
    \includegraphics[width=0.85\textwidth]{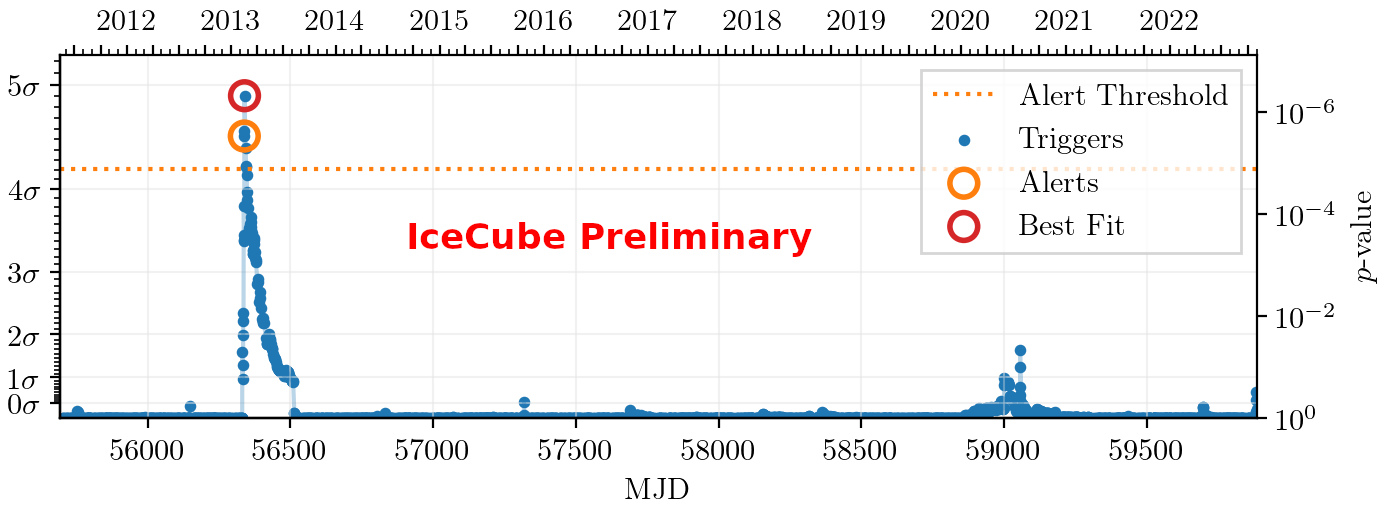}
    \caption{The pre-trial significances of the best fit flare time-window as a function of the event trigger time for the hottest all-sky source location over 11.5 years of data.
    The orange circled point represents where an alert would have been sent if the alerts had been active in 2013.
    The red circle indicates the peak in significance found for the source which after trials corrections drops in signficance to a p-value of 0.482.}
    \label{fig:allsky_hotspot}
\end{figure}
\begin{figure}
    \centering
    \includegraphics[width=0.55\textwidth]{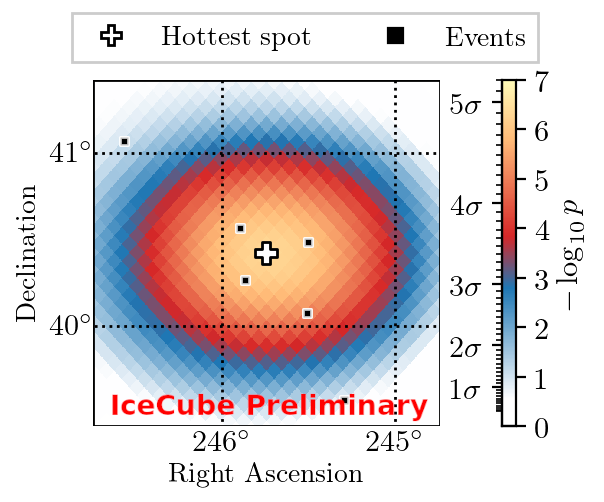}
    \caption{The localization of the most significant all-sky neutrino flare. The color scale represents the pre-trial p-values at the time of the hottest spot trigger event in the region around the hottest spot location (white cross).
    The black squares represent the best-fit locations of the contributing events.}
    \label{fig:hotspot_skymap}
\end{figure}
\begin{figure}
    \centering
    \includegraphics[width=0.8\textwidth]{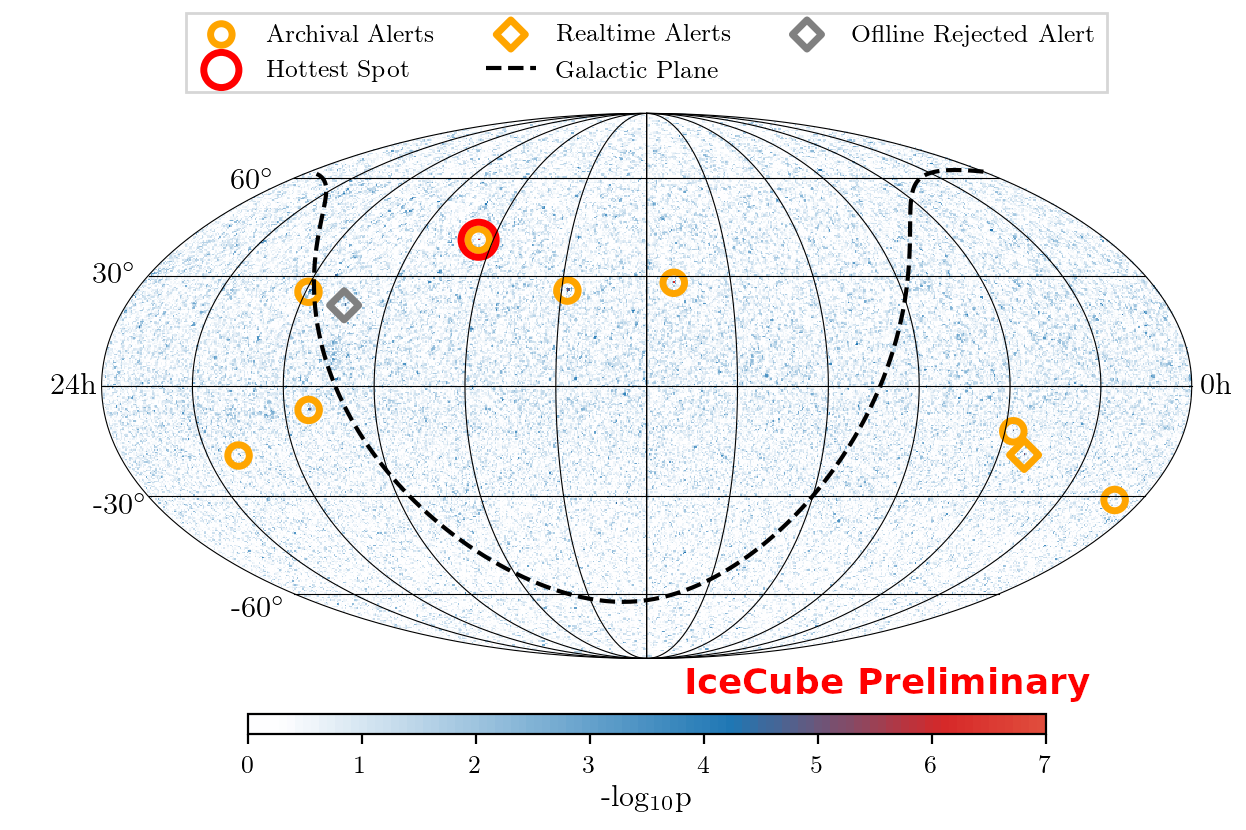}
    \caption{Figure showing the pre-trial p-values for the most significant flare at each pixel in the sky. The orange circles represent the locations where the p-value is greater than the all-sky alert threshold, but occurred in the archival data before the alert stream was active so no alert was sent. The diamonds represent alerts which were sent to the community. The orange diamond was confirmed by the archival data; whereas, the grey diamond was an alert which was not found in the archival data because some of the contributing events occurred during detector runs that did not pass offline quality cuts. The red circle represents the hottest flare location.}
    \label{fig:all_sky_results}
\end{figure}

The most significant spot from the all-sky analysis was found in the northern sky at a declination of 40.42°.
It also occurred in the archival data before the alert stream was active. 
The pre-trial significance was found to be 4.90$\sigma$; however, after correcting for trials, the p-value amounts to 0.482.
The duration of the best-fit time window was 9.4 days which was found in 2013 as shown in Figure~\ref{fig:allsky_hotspot}.
The localization of the best-fit flare is shown in Figure~\ref{fig:hotspot_skymap}, where the pre-trial p-value is calculated for all pixels in the region of the hottest spot at the time of the most significant trigger.
These results highlight the importance of a multimessenger approach since searching through the IceCube data alone results in a large number of trials, but coincident very-high-energy gamma-ray emission could boost the significance.
The locations of the archival and realtime alerts are circled in Figure~\ref{fig:all_sky_results}.
Neutrino flare curves for each source in the source-list analysis and each archival and realtime all-sky alert location, like those shown in Figures~\ref{fig:source_hotspot} and~\ref{fig:allsky_hotspot}, will be provided in a future publication and data release.

\section{Conclusions and Outlook}\label{con}
The alert stream presented here looks for clusters of neutrinos in location and time to provide targets of opportunity for imaging air Cherenkov telescopes (IACTs).
This optimizes the synergy between IceCube and IACTs by using IceCube's ability to observe the whole sky with a duty cycle of around 99\% to trigger follow-up by the sensitive IACTs.
An offline search found no statistically significant flares using the IceCube data alone; however, a multimessenger analysis could provide the statistical strength as background coincidence between a neutrino flare and very-high-energy (VHE) gamma-rays is less likely~\cite{IceCube:2023VHEGFU, HESS:GFUFollowUp}.

The alert streams presented here have the possibility for future improvements.
Advancements in IceCube's event selection and event reconstruction techniques can improve our sensitivity to signal neutrino flares.
Also, within the past 10 years of operation, IceCube has increased our knowledge about astrophysical neutrinos and their possible sources. 
While looking for correlation of neutrino flares with VHE gamma-rays is a logical first hypothesis, it has been shown that looking for correlations with other wavelengths of light might be more fruitful.
We plan in the future to generate a public version of the alert stream with a broader source list such that followup can be done by other types of telescopes.
Finally, as more neutrino telescopes are put into operation, collaboration between neutrino telescopes could be advantageous especially in the realtime domain.

% Bibtex references:
\bibliographystyle{ICRC}
\bibliography{references}

% Alternatively, you can include references by hand:
%\begin{thebibliography}{99}
%\bibitem{...}
%
%\end{thebibliography}

\clearpage

%The following list of authors, affiliations and funding agencies will be updated at the day of submission. The following template is a placeholder generated via https://authorlist.icecube.wisc.edu/icecube on March 25, 2023 and will be updated.
\section*{Full Author List: IceCube Collaboration}

\scriptsize
\noindent
R. Abbasi$^{17}$,
M. Ackermann$^{63}$,
J. Adams$^{18}$,
S. K. Agarwalla$^{40,\: 64}$,
J. A. Aguilar$^{12}$,
M. Ahlers$^{22}$,
J.M. Alameddine$^{23}$,
N. M. Amin$^{44}$,
K. Andeen$^{42}$,
G. Anton$^{26}$
% ...
\\
\\
% ...
$^{12}$ Universit{\'e} Libre de Bruxelles, Science Faculty CP230, B-1050 Brussels, Belgium \\
% ...
$^{17}$ Department of Physics, Loyola University Chicago, Chicago, IL 60660, USA \\
$^{18}$ Dept. of Physics and Astronomy, University of Canterbury, Private Bag 4800, Christchurch, New Zealand \\
% ...
$^{22}$ Niels Bohr Institute, University of Copenhagen, DK-2100 Copenhagen, Denmark \\
$^{23}$ Dept. of Physics, TU Dortmund University, D-44221 Dortmund, Germany \\
% ...
$^{26}$ Erlangen Centre for Astroparticle Physics, Friedrich-Alexander-Universit{\"a}t Erlangen-N{\"u}rnberg, D-91058 Erlangen, Germany \\
% ...
$^{40}$ Dept. of Physics and Wisconsin IceCube Particle Astrophysics Center, University of Wisconsin{\textendash}Madison, Madison, WI 53706, USA \\
% ...
$^{42}$ Department of Physics, Marquette University, Milwaukee, WI, 53201, USA \\
% ...
$^{44}$ Bartol Research Institute and Dept. of Physics and Astronomy, University of Delaware, Newark, DE 19716, USA \\
% ...
$^{63}$ Deutsches Elektronen-Synchrotron DESY, Platanenallee 6, 15738 Zeuthen, Germany  \\
$^{64}$ Institute of Physics, Sachivalaya Marg, Sainik School Post, Bhubaneswar 751005, India \\
% ...

\subsection*{Acknowledgements}

\noindent
The authors gratefully acknowledge the support from the following agencies and institutions:
USA {\textendash} U.S. National Science Foundation-Office of Polar Programs,
U.S. National Science Foundation-Physics Division,
U.S. National Science Foundation-EPSCoR,
Wisconsin Alumni Research Foundation,
Center for High Throughput Computing (CHTC) at the University of Wisconsin{\textendash}Madison,
Open Science Grid (OSG),
Advanced Cyberinfrastructure Coordination Ecosystem: Services {\&} Support (ACCESS),
Frontera computing project at the Texas Advanced Computing Center,
U.S. Department of Energy-National Energy Research Scientific Computing Center,
Particle astrophysics research computing center at the University of Maryland,
Institute for Cyber-Enabled Research at Michigan State University,
and Astroparticle physics computational facility at Marquette University;
Belgium {\textendash} Funds for Scientific Research (FRS-FNRS and FWO),
FWO Odysseus and Big Science programmes,
and Belgian Federal Science Policy Office (Belspo);
Germany {\textendash} Bundesministerium f{\"u}r Bildung und Forschung (BMBF),
Deutsche Forschungsgemeinschaft (DFG),
Helmholtz Alliance for Astroparticle Physics (HAP),
Initiative and Networking Fund of the Helmholtz Association,
Deutsches Elektronen Synchrotron (DESY),
and High Performance Computing cluster of the RWTH Aachen;
Sweden {\textendash} Swedish Research Council,
Swedish Polar Research Secretariat,
Swedish National Infrastructure for Computing (SNIC),
and Knut and Alice Wallenberg Foundation;
European Union {\textendash} EGI Advanced Computing for research;
Australia {\textendash} Australian Research Council;
Canada {\textendash} Natural Sciences and Engineering Research Council of Canada,
Calcul Qu{\'e}bec, Compute Ontario, Canada Foundation for Innovation, WestGrid, and Compute Canada;
Denmark {\textendash} Villum Fonden, Carlsberg Foundation, and European Commission;
New Zealand {\textendash} Marsden Fund;
Japan {\textendash} Japan Society for Promotion of Science (JSPS)
and Institute for Global Prominent Research (IGPR) of Chiba University;
Korea {\textendash} National Research Foundation of Korea (NRF);
Switzerland {\textendash} Swiss National Science Foundation (SNSF);
United Kingdom {\textendash} Department of Physics, University of Oxford.

\end{document}